\documentclass[aps,pra,epsfigure,twocolumn]{revtex4}
\usepackage{dcolumn}    
\usepackage{bm} 
\usepackage{graphicx}
\usepackage{amsmath}    
\usepackage{latexsym}
\usepackage{amsfonts}   
\usepackage{amssymb}
\usepackage{array}      
\usepackage{epsfig}
\usepackage{txfonts}
\usepackage[colorlinks=true,linkcolor=blue,urlcolor=blue,citecolor=blue]{hyperref}
\usepackage{color}
\usepackage{hyperref}

\newcommand{\ket}[1]{\left\vert#1\right\rangle}

\begin{document}
 \title{Dynamics and asymptotics of correlations in a many-body localized system}
\author{Steve Campbell,$^{1,2}$ Matthew J. M. Power,$^1$ and Gabriele De Chiara$^1$}
\affiliation{
$^1$\mbox{Centre for Theoretical Atomic, Molecular and Optical Physics, Queen's University Belfast, Belfast BT7 1NN, United Kingdom}\\
$^2$Istituto Nazionale di Fisica Nucleare, Sezione di Milano, \& Dipartimento di Fisica, Universit{\`a} degli Studi di Milano, Via Celoria 16, 20133 Milan, Italy
}

\begin{abstract}
We examine the dynamics of nearest-neighbor bipartite concurrence and total correlations in the spin-1/2 $XXZ$ model with random fields. We show, starting from factorized random initial states, that the concurrence can suffer entanglement sudden death in the long time limit and therefore may not be a useful indicator of the properties of the system. In contrast, we show that the total correlations capture the dynamics more succinctly, and further reveal a fundamental difference in the dynamics governed by the ergodic versus many-body localized phases, with the latter exhibiting dynamical oscillations. Finally, we consider an initial state composed of several singlet pairs and show that by fixing the correlation properties, while the dynamics do not reveal noticeable differences between the phases, the long-time values of the correlation measures appear to indicate the critical region.
\end{abstract}
\date{\today}
\maketitle

\section{Introduction}
The study of entanglement in strongly correlated systems focused initially on the critical properties of the ground state of one-dimensional spin chains close to a quantum phase transition~\cite{OsterlohNature,Osborne} (see Ref.~\cite{FazioRMP} for a comprehensive review). After these early studies, much interest has been devoted to the study of the time evolution of entanglement after a sudden quench or a continuous change of the Hamiltonian~\cite{Plastina,CC,DeChiaraJSTAT}. Of particular interest are those works dealing with disordered systems and the possibility of inducing, in the presence of interactions, many-body localisation (MBL)~\cite{Basko}, see the recent reviews Refs.~\cite{nandkishore2015many,MooreReview,LaflorencieReview} and references therein. In contrast to Anderson localization, in MBL systems in one dimension, localization does not occur for an infinitesimal disorder but for a non-zero value.

In the last few years, interest on the MBL phase has grown remarkably fast. It is now well established that this phase is characterised by the absence of thermalisation, notwithstanding the presence of interactions, due to the emergence of local conservation laws similarly to integrable systems. Such considerations have helped to develop useful tools for studying the MBL phase using local probes~\cite{Serbyn1,Serbyn2,Serbyn3,Huse4}. Recent studies have also shown the use of quantities such as quantum mutual information and entanglement are useful for examining the transition to the MBL phase~\cite{PollmanPRL,BeraPRB,GooldPRB,Sarandy}. Energy eigenstates in the middle of the spectrum of an MBL Hamiltonian fulfil the entanglement area-law and gives rise to a slow logarithmic growth of entanglement after a sudden quench. MBL has been recently observed in experiments with ultracold atoms \cite{Schreiber,Kondov,Choi} and trapped ions \cite{Smith}. 

While often block entanglement entropy is the focus, in this work we consider the dynamical onset of the MBL phase and study the dynamics of the nearest-neighbor concurrence, a faithful measure of two-spin entanglement, and the total correlations, which measure all the correlations, classical and quantum, shared by all the spins in the chain. Although these two quantities have been analysed for the centre of the spectrum of an interacting many-body Hamiltonian \cite{BeraPRB,GooldPRB} (see also Ref.~\cite{Sarandy} where more general pairwise correlations are considered), the study of the evolution of these quantities and the corresponding asymptotic properties is still missing.

To this end and following Ref.~\cite{HusePRB}, we fix the initial state rather than focusing on a particular energy band in the spectrum~\cite{BeraPRB,GooldPRB,Sarandy,AletPRB}. To begin we will focus on a random, pure, separable state analogous to the situation in Ref.~\cite{HusePRB}. Our initial states thus uniformly sample the full spectrum of the system, i.e. the initial energy distribution forms a Gaussian centred around zero. This is equivalent to exploring a high temperature region of the energy spectrum. We also consider an initial state composed of tensor products of singlets. This state, similarly to MBL states, is locally entangled but does not have long range entanglement. This ensures the state initially has entanglement localized between certain spin pairs and fixes the marginal probability distributions. As we will see both settings reveal interesting features of the nature of the ergodic-MBL transition.

\section{Preliminaries}
We consider the spin-1/2 $XXZ$ model with periodic boundary conditions subject to random disorder (longitudinal fields), $h_i$, applied to each spin. The Hamiltonian is given by
\begin{equation}
\mathcal{H} = \frac{1}{2} \sum_{i=1}^{L-1} \left( \sigma_x^{i}\otimes \sigma_x^{i+1} +\sigma_y^{i}\otimes \sigma_y^{i+1} + \Delta \sigma_z^{i}\otimes \sigma_z^{i+1} \right) + \sum_{i=1}^{L} h_i \sigma_z^i.
\end{equation}
The random fields $h_i$ are uniformly chosen from the interval $\left[ -\eta, \eta \right]$. For $\Delta > 0$ this model will exhibit a transition between an ergodic and an MBL phase~\cite{DeltaComment,EnssArXiv}, which is dependent on the magnitude of the disorder strength $\eta$. In what follows we will consider $\Delta=1$, unless otherwise stated. For this interaction strength the current best estimates for the critical disorder strength is predicted to occur at $\eta_c \approx 3.7$ (although some estimates can be as low as $\eta_c\approx 3.5$), determined using energy resolved calculations~\cite{AletPRB} and total correlations of the diagonal ensemble~\cite{GooldPRB}. However there are evidences of an extended though not-ergodic phase for $\eta<\eta_c$ \cite{GooldPRB,DeLuca}. We remark that a recent study has shown a closely related model where the system is quasi-periodic rather than random appears to be in a distinct universality class~\cite{Huse5}.

We will focus on two figures of merit in particular: the concurrence and the total correlations. Concurrence is a measure of entanglement valid for arbitrary states of two qubits. It is defined in terms of the eigenvalues $\lambda_1\,{\geq}\, \lambda_{2,3,4}$ of the spin-flipped density matrix $\rho_{12}(\sigma_y\otimes\sigma_y)\rho^*_{12} (\sigma_y\otimes\sigma_y)$ as
\begin{equation}
C=\text{max}\left[0,\sqrt{\lambda_1}-\sum_{i=2}^{4}\sqrt{\lambda_i}\right].
\end{equation}
Therefore, when evaluating the entanglement we will focus on the first two spins of the chain, i.e. $\rho_{12} = \text{Tr}_{i\neq1,2} \left[ \rho \right]$. We remark the relationship between concurrence and MBL was recently explored in the high energy region of the spectrum~\cite{BeraPRB}.

Total correlations are defined as the information shared between all constituents of the state. As such, and unlike entanglement, the total correlation encompasses both classical and quantum natures. We define the total correlations as
\begin{equation}
\label{totalcorr}
I = \sum_i S(\rho_i) - S(\rho)
\end{equation}
where $S(\cdot)$ denotes the von Neumann entropy, $\rho$ is the total density matrix of the system and $\rho_i$ is the reduced density matrix of spin $i$. For two-spins this is equivalent to the mutual information shared between them. Furthermore, since in what follows the state is always pure and therefore $S(\rho) = 0$, $I$ is simply the sum of the von Neumann entropy of the marginals. In Ref.~\cite{GooldPRB} this figure of merit was used to explore the ergodic-MBL transition, again in the high energy region.

\begin{figure}[t]
{\bf (a)}\hskip0.45\columnwidth {\bf (d)}
\includegraphics[width=0.5\columnwidth]{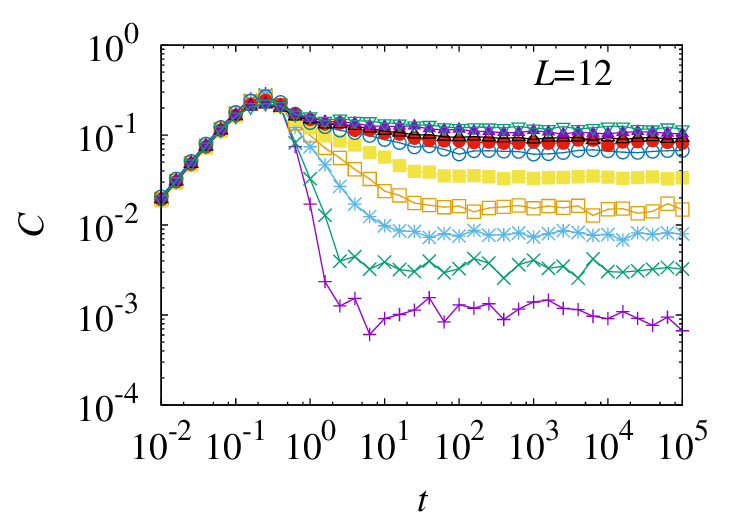}\includegraphics[width=0.5\columnwidth]{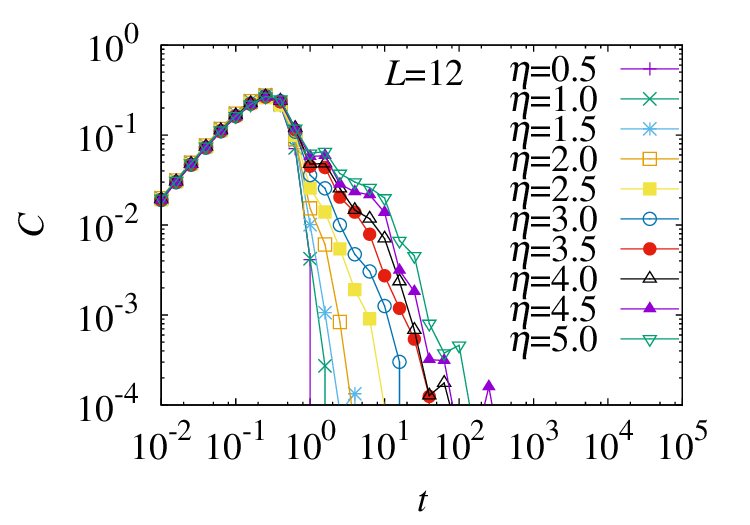}
{\bf (b)}\hskip0.45\columnwidth {\bf (e)}
\includegraphics[width=0.5\columnwidth]{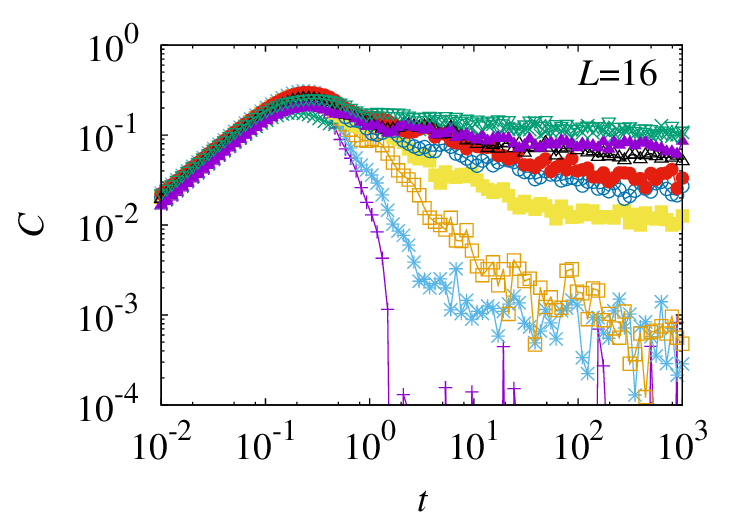}\includegraphics[width=0.5\columnwidth]{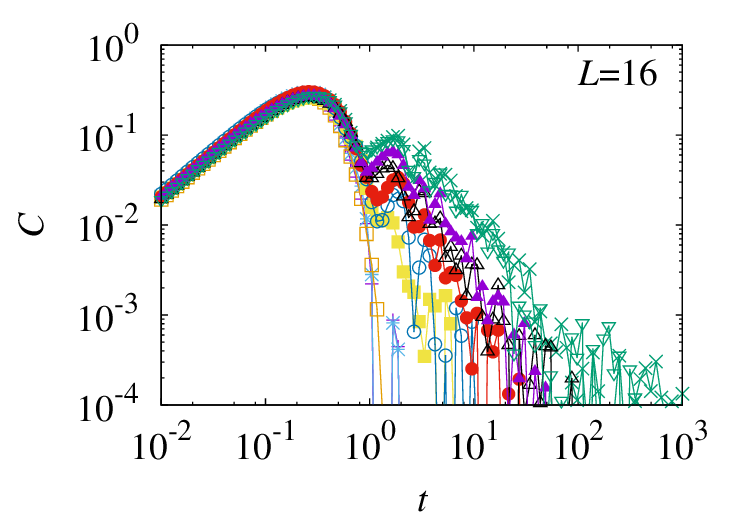}
{\bf (c)}\hskip0.45\columnwidth {\bf (f)}
\includegraphics[width=0.5\columnwidth]{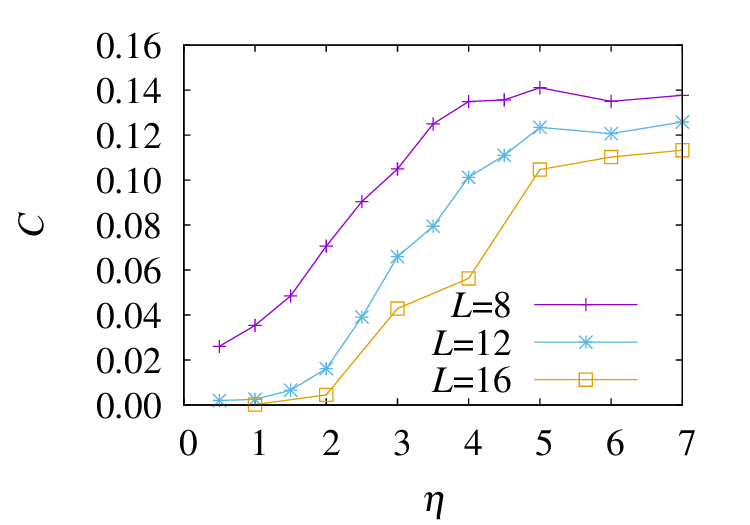}\includegraphics[width=0.5\columnwidth]{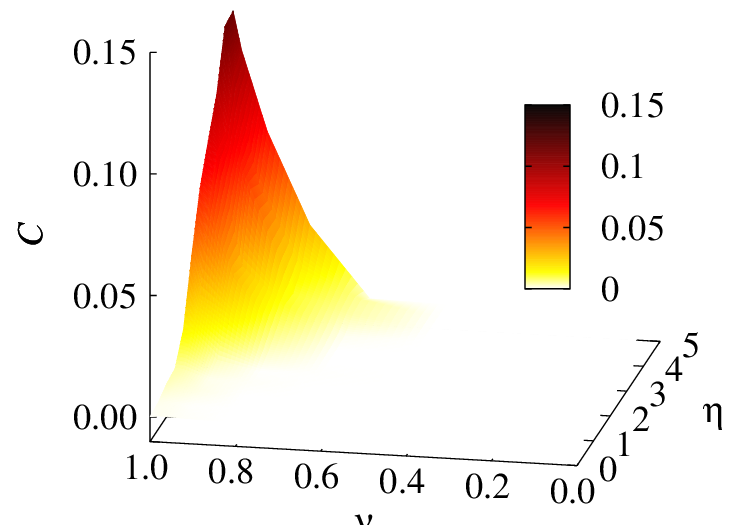}
\caption{(Color online) {\bf (a-b-d-e)} Dynamics of nearest neighbor concurrence fixing $\Delta=1$ for the disorder strength, $\eta=0.5$ (lowest, purple crosses) to 5 (top-most, green crosses) in steps of 0.5. In the left column we fix $v=1$ with {\bf (a)} $L=12$ and {\bf (b)} $L=16$. In the right column we fix $v=0.5$ with {\bf (d)} $L=12$ and {\bf (e)} $L=16$. {\bf (c)} Asymptotic value of the nearest neighbor concurrence against disorder, $\eta$, for $v=1$. {\bf (f)} Asymptotic value of the nearest neighbor concurrence against disorder, $\eta$, and $v$ for $L=12$. The color-coding for the panels {\bf (a-b-e)} is the same as in panel {\bf (d)}.}
\label{fig1}
\end{figure}
\section{Results}
\subsection{Random initial states}
As an initial state, each spin at site $i$ is prepared in a pure state
\begin{equation}
\ket{\psi_i} = \cos \left(\frac{\theta_i}{2} \right) \ket{0_i} + e^{i \phi_i} \sin\left(\frac{\theta_i}{2} \right) \ket{1_i},
\label{eq:psii}
\end{equation}
where $\cos(\theta_i)$ is chosen randomly to be $\pm v$ and $\phi_i$ is chosen randomly from $[ 0, 2\pi)$, see Ref.~\cite{HusePRB} Fig. 1. We remark this sampling means that each spin is at a fixed angle above or below the equatorial plane of the Bloch sphere, pointing in a random direction and therefore we are not considering random states in the typical sense sampled according to the Haar measure. Our initial state is then 
\begin{equation}
\label{randompsi}
\ket{\psi} = \bigotimes_{i=1}^L \ket{\psi_i}. 
\end{equation}
We evolve this state for many realizations of the disorder, with each one starting from a different $\ket{\psi}$. We perform at least 1000 simulations for each value of $\eta$ in order to ensure good convergence. 

In Fig.~\ref{fig1} we examine the dynamics of the nearest neighbor concurrence. In panels {\bf (a)} and {\bf (b)} we take $v=1$, this corresponds to the situation in which each individual spin, i.e. its Bloch vector, in Eq.~\eqref{randompsi} is randomly chosen to point along the $\pm~z$-axis. We see the initial dynamics are insensitive to the magnitude of the disorder. However, after $t\sim 10^{-0.5}$ and as $\eta$ is increased, the amount of nearest neighbor concurrence is also increased. Furthermore, small $\eta$ witnesses a sharp drop in the amount of entanglement shared between the two spins before settling into its long-time value, while for larger values of $\eta$, entering the MBL phase, the system takes longer to settle. Such a behavior is consistent with the slow growth of block entropy~\cite{HusePRB}. Comparing panels {\bf (a)} and {\bf (b)} in Fig.~\ref{fig1} we see that qualitatively these features persist regardless of the system size $L$. Panel {\bf (c)} shows the asymptotic values for $L=8,~12$ and $16$. These results are in agreement with those reported in Ref.~\cite{BeraPRB} where the entanglement properties of states in the middle of the spectrum of the Hamiltonian were examined, similarly showing that the (total) nearest neighbor concurrence grows from zero in the ergodic phase to comparatively large values when the system transitions into the MBL phase. 

\begin{figure}[t]
\centering
{\bf (a)}\hskip0.45\columnwidth {\bf (d)}
\includegraphics[width=0.5\columnwidth]{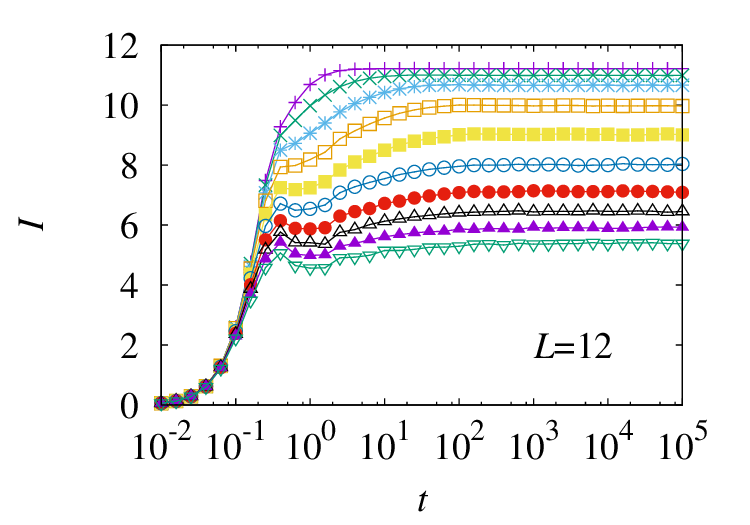}\includegraphics[width=0.5\columnwidth]{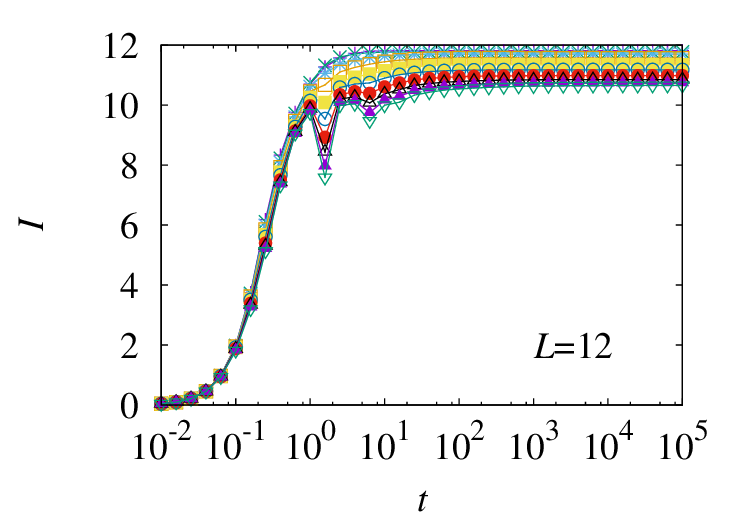}
{\bf (b)}\hskip0.45\columnwidth {\bf (e)}
\includegraphics[width=0.5\columnwidth]{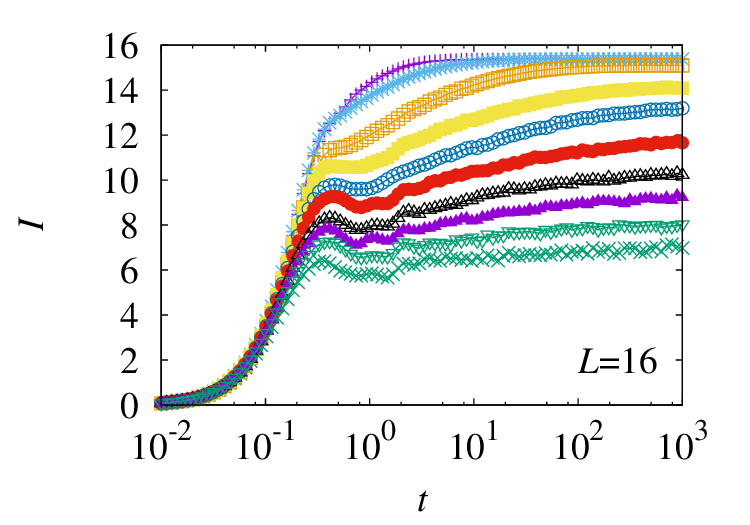}\includegraphics[width=0.5\columnwidth]{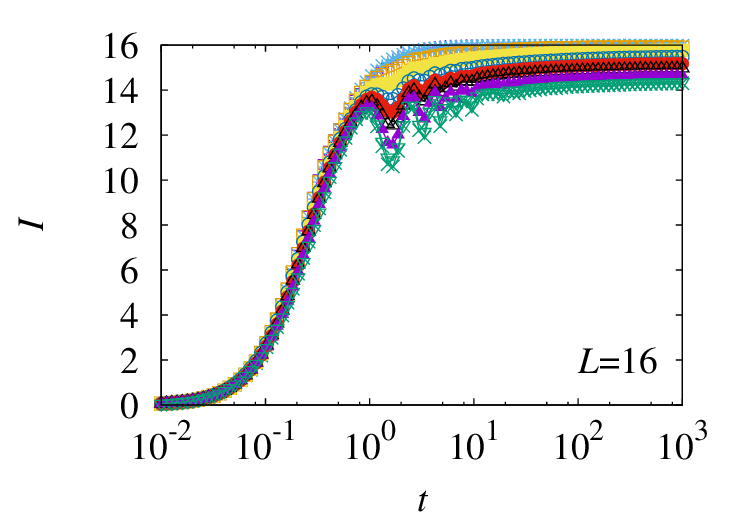}
{\bf (c)}\hskip0.45\columnwidth {\bf (f)}
\includegraphics[width=0.5\columnwidth]{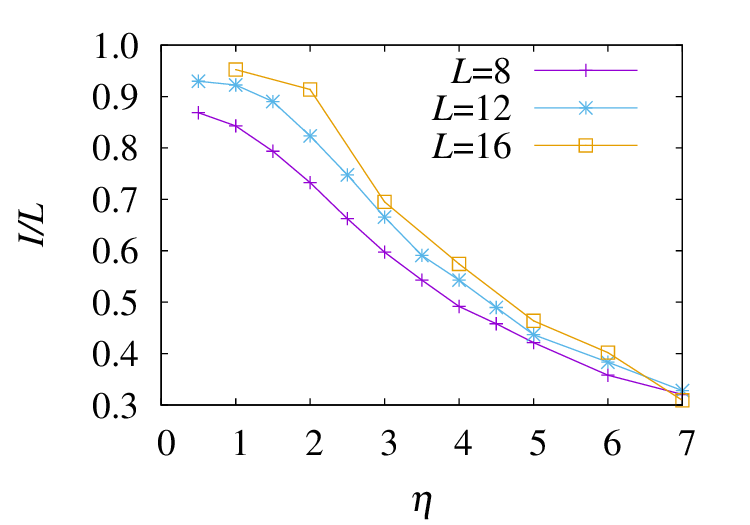}\includegraphics[width=0.5\columnwidth]{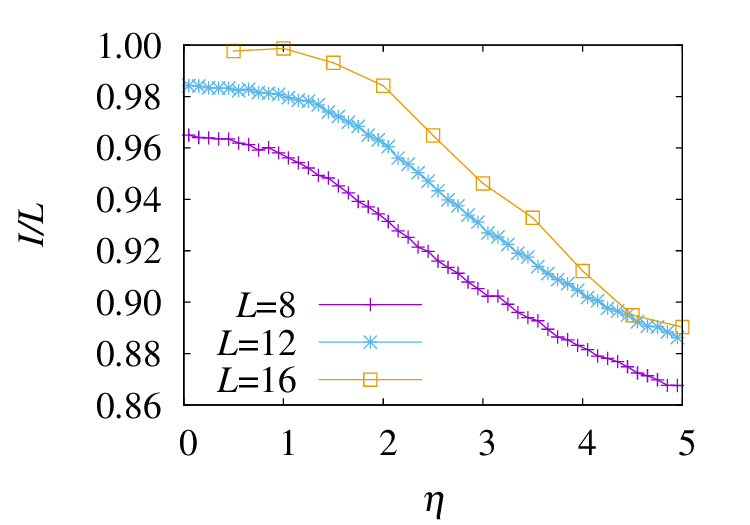}
\caption{(Color online) {\bf (a)} - {\bf (c)} Dynamics of the total correlations fixing $\Delta=1$ for the disorder strength, $\eta=0.5$ to 5 (as in Fig.~\ref{fig1}). In the left column we fix $v=1$ with {\bf (a)} $L=12$ and {\bf (b)} $L=16$. In the right column we fix $v=0.5$ with {\bf (d)} $L=12$ and {\bf (e)} $L=16$. Also shown is the asymptotic value of the total correlations (rescaled with $L$) against disorder, $\eta$, for {\bf (c)} $v=1$ and {\bf (f)} $v=0.5$. The color-coding for the panels {\bf (a-b-d-e)} is the same as in Fig.~\ref{fig1}.}
\label{fig2}
\end{figure}

We perform the same simulations only altering the initial state such that $v=0.5$. In this case the short-time dynamics are qualitatively the same as before. Once again, while initially all values of $\eta$ present the same dynamics, as we increase the disorder strength the systems evolving within the MBL phase settle slower than in the ergodic phase. However, an important difference arises: now the asymptotic value of the nearest neighbor concurrence tends to zero. Panel {\bf (f)} shows that the long-time behaviour of the concurrence is strongly affected by the choice of  initial state. Taking $v=1$ we see that the concurrence is sensitive to the disorder strength precisely in line with Ref.~\cite{BeraPRB}, however for other values the concurrence quickly reaches zero, a phenomenon known as entanglement sudden death (ESD)~\cite{ESD}. This may naively lead one to assume that the long-time entanglement is largely unaffected by the disorder strength. Indeed, such a behaviour was reported in Ref.~\cite{HusePRB} when studying the block entropy and examining values of $v\in(0,0.84)$ (although we remark this study focused on the MBL phase with $\eta\geq 6$). However, as shown in Fig.~\ref{fig1} {\bf (f)} this range corresponds to nearest-neighbor concurrence being zero, and therefore the invariance reported maybe due to such pathological features. 

We recall that the approach employed here is expected to model the high temperature behaviour in the long-time limit. Similarly, directly accessing the middle of the spectrum, as done in Ref.~\cite{BeraPRB}, is also expected to reproduce the same high temperature features. We have checked that the initial energy distribution obtained by taking the class of states \eqref{eq:psii} is qualitatively similar (although generally broader) to the one obtained by taking a few tens of states in the middle of the spectrum as in Ref.~\cite{BeraPRB}. Here, we have shown that great care must be taken when considering entanglement measures such as the concurrence. Changing the initial state can lead to seemingly contradictory conclusions, stemming from the (in this case) pathological occurrence of ESD. We therefore seek to employ a different figure of merit to alleviate this problem.

In Fig.~\ref{fig2} we examine the dynamics and asymptotic values for the total correlations, Eq.~\eqref{totalcorr}. While all of the main qualitative features persist we can now more clearly identify the role $v$ plays. In panels {\bf (a-b-c)}, $v=1$ and again the state is very sensitive to the disorder strength $\eta$, with the total correlations decreasing as $\eta$ grows. This is in agreement with Fig.~\ref{fig1} {\bf (b)} and {\bf (c)} for the same`favorable' value of $v$, i.e. one that maintains a non-zero value of concurrence in the long time limit and thus does not exhibit ESD. We remark that it is intuitive that a measure of total correlations, that encompasses all classical and quantum aspects, should decrease when  the bipartite entanglement grows since, due to the monogamy properties of the entanglement, larger bipartite entanglement generally necessitates a reduction in the the total quantum correlations. In panels {\bf (d-e-f)} we fix $v=0.5$. In this case a remarkable feature emerges that was not immediately evident when studying the concurrence. In the ergodic phase $I$ grows monotonically until settling to its long time value. As the disorder is increased, and we enter the MBL phase, we see the emergence of oscillations. These oscillations persist for a significant time, gradually dissipating until the system settles to its asymptotic value. Indeed, in the MBL phase we expect the system to store some memory of the initial state in the dynamics, and this would appear to be evidenced by these oscillations. Additionally, in panel {\bf (f)} we now see that the asymptotic values for the total correlations reflect the changes in the disorder strength. Furthermore, the magnitude of this effect is significantly larger than compared to block entropy~\cite{HusePRB}.

\begin{figure}[t]
\centering
{\bf (a)}\hskip0.45\columnwidth {\bf (b)} 
\includegraphics[width=0.5\columnwidth]{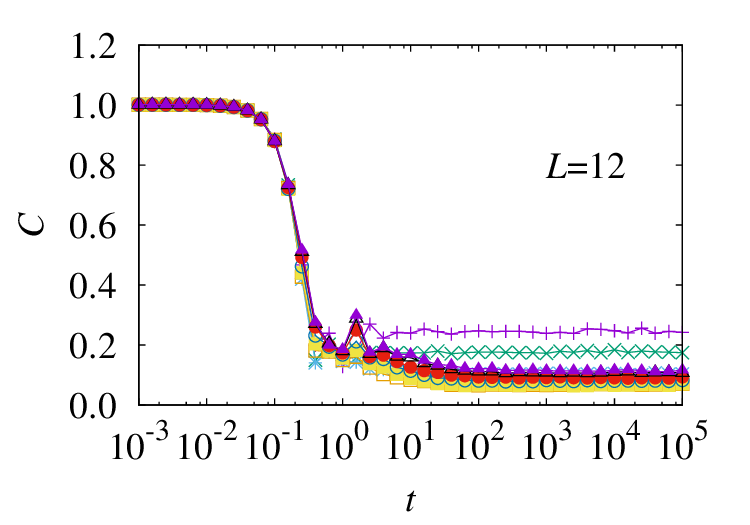}
\hspace{-0.4cm}
\includegraphics[width=0.5\columnwidth]{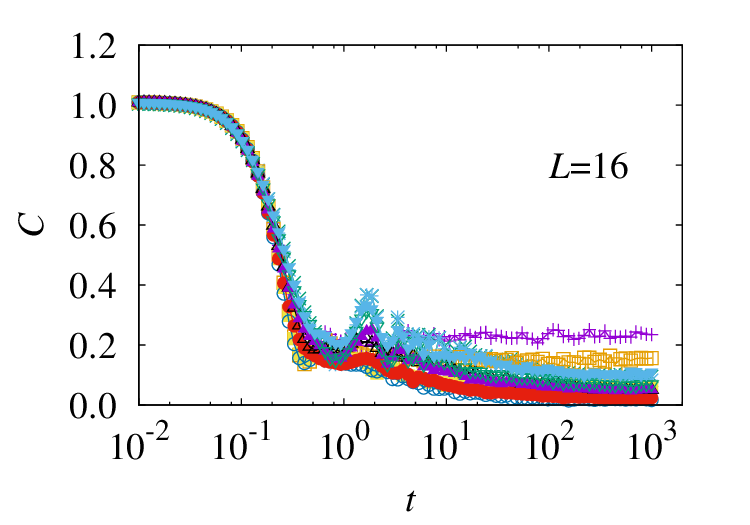}
{\bf (c)}\hskip0.45\columnwidth {\bf (d)}
\includegraphics[width=0.5\columnwidth]{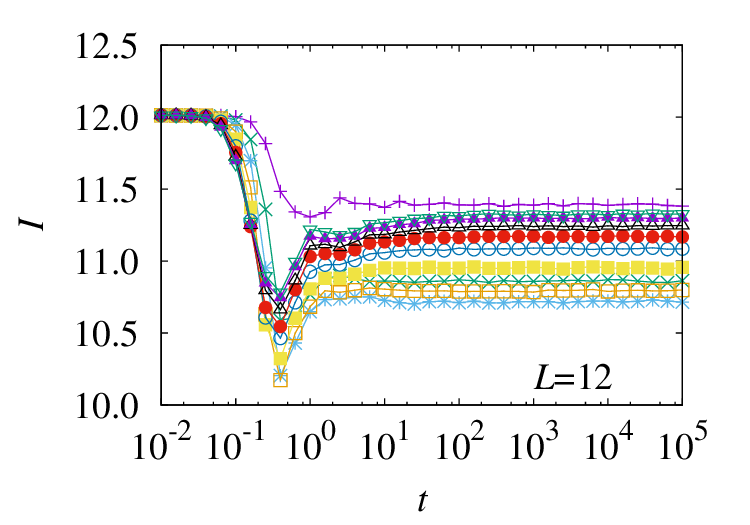}
\hspace{-0.4cm}
 \includegraphics[width=0.5\columnwidth]{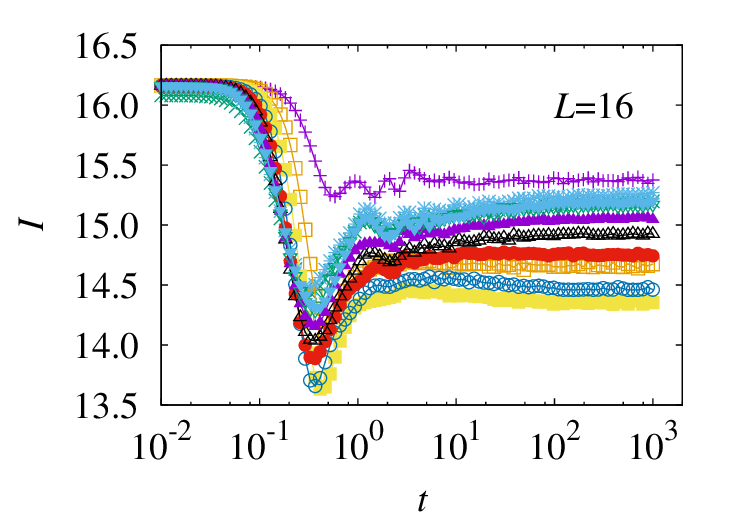}
\caption{(Color online) {\bf (a-b)}: Dynamics of nearest neighbour concurrence and {\bf (c-d)}: dynamics of total correlations for the initial state composed of singlet pairs, Eq.~\eqref{singlets}. We fix $\Delta=1$ and take  values of $\eta\in[0.5,5]$ in steps of 0.5. {\bf (a-c)} $L=12$ and {\bf (b-d)} $L=16$. The color-coding for the panels is the same as in panel Fig.~\ref{fig1}~{\bf (a)}. }
\label{fig34}
\end{figure}

\subsection{Singlet pairs}
We next turn our attention to a different initial configuration for the chain. Starting from a singlet $\ket{\Psi^-_{i}} = \tfrac{1}{\sqrt{2}} (\ket{01} - \ket{10})_{(2i-1),2i}$, we take our initial state to be
\begin{equation}
\label{singlets}
\ket{\psi} = \bigotimes_{i=1}^{L/2} \ket{\Psi^-_i}.
\end{equation}
The initial energy is now fixed regardless of the random fields $h_i$ and it is negative. We evolve Eq.~\eqref{singlets} in precisely the same manner as done previously and in what follows we study the dynamical properties and asymptotic values of our figures of merit.

\begin{figure}[t]
\centering
{\bf (a)}\hskip0.45\columnwidth {\bf (b)}\\ 
\includegraphics[width=0.5\columnwidth]{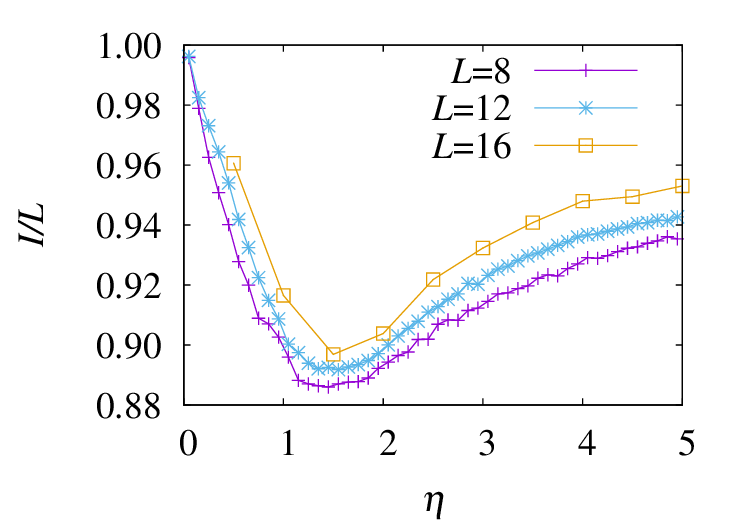}\includegraphics[width=0.5\columnwidth]{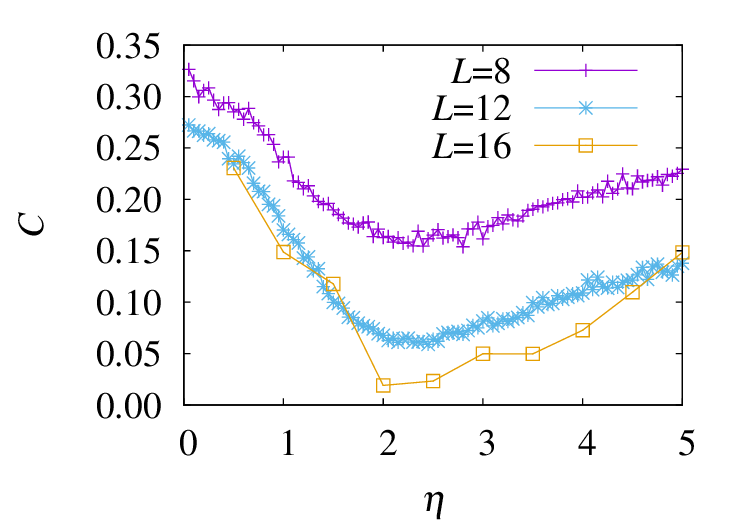}
\caption{(Color online) Asymptotic value of the {\bf (a)} total correlations (rescaled with $L$) and {\bf (b)} nearest neighbor concurrence against the disorder strength $\eta$.}
\label{fig5}
\end{figure}

In Fig.~\ref{fig34}  we show the concurrence and total correlations. For both, the initially localized correlation is frozen for a short time window, which is then followed by an exponential decay with minimal dynamical fluctuations before settling to a non-zero long-time value, a trend followed regardless of the length considered and consistent for all disorder strengths. Here we find that the long-time value of both quantifiers is strongly affected by the disorder strength. In Fig.~\ref{fig5} we explore the behavior of these asymptotic values against disorder strength more closely. Panel {\bf (a)} shows the total correlations, rescaled by $L$, for several chain lengths. We see the curves remain close to one another for increasing lengths, confirming the extensive nature of $I$. Additionally, notice that the total correlations {\it increase} slightly as we increase system size. The concurrence shown in panel {\bf (b)} exhibits a similar qualitative behavior, however with one important difference, we now see that the asymptotic value {\it decreases} as the system is enlarged. Interestingly, as we increase $\eta$ moving from the ergodic to MBL phases, this value decreases. In the MBL phase, it reaches a minimum value after which it starts to grow to its large $\eta$ value.

\begin{figure}[t]
\centering
{\bf (a)}\hskip0.45\columnwidth {\bf (b)}
\includegraphics[width=0.5\columnwidth]{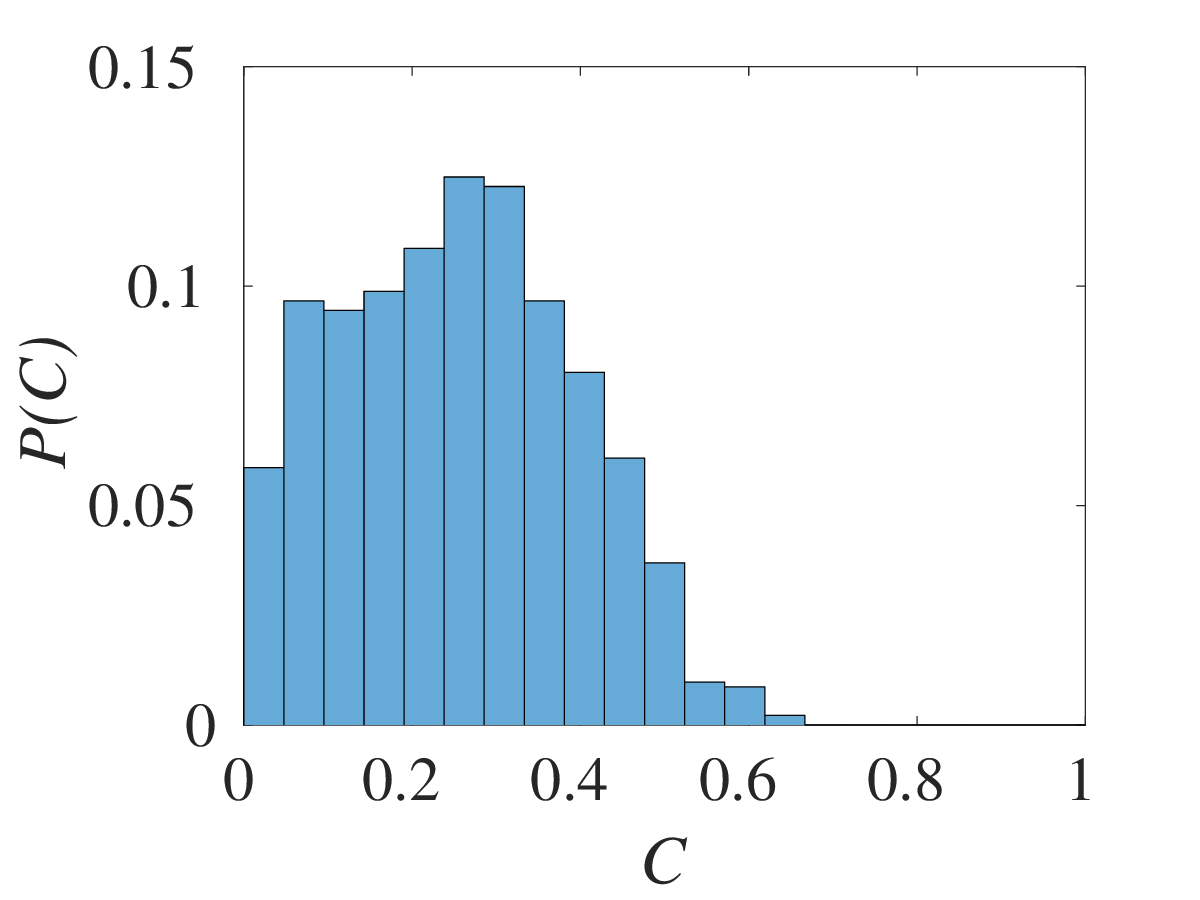}\includegraphics[width=0.5\columnwidth]{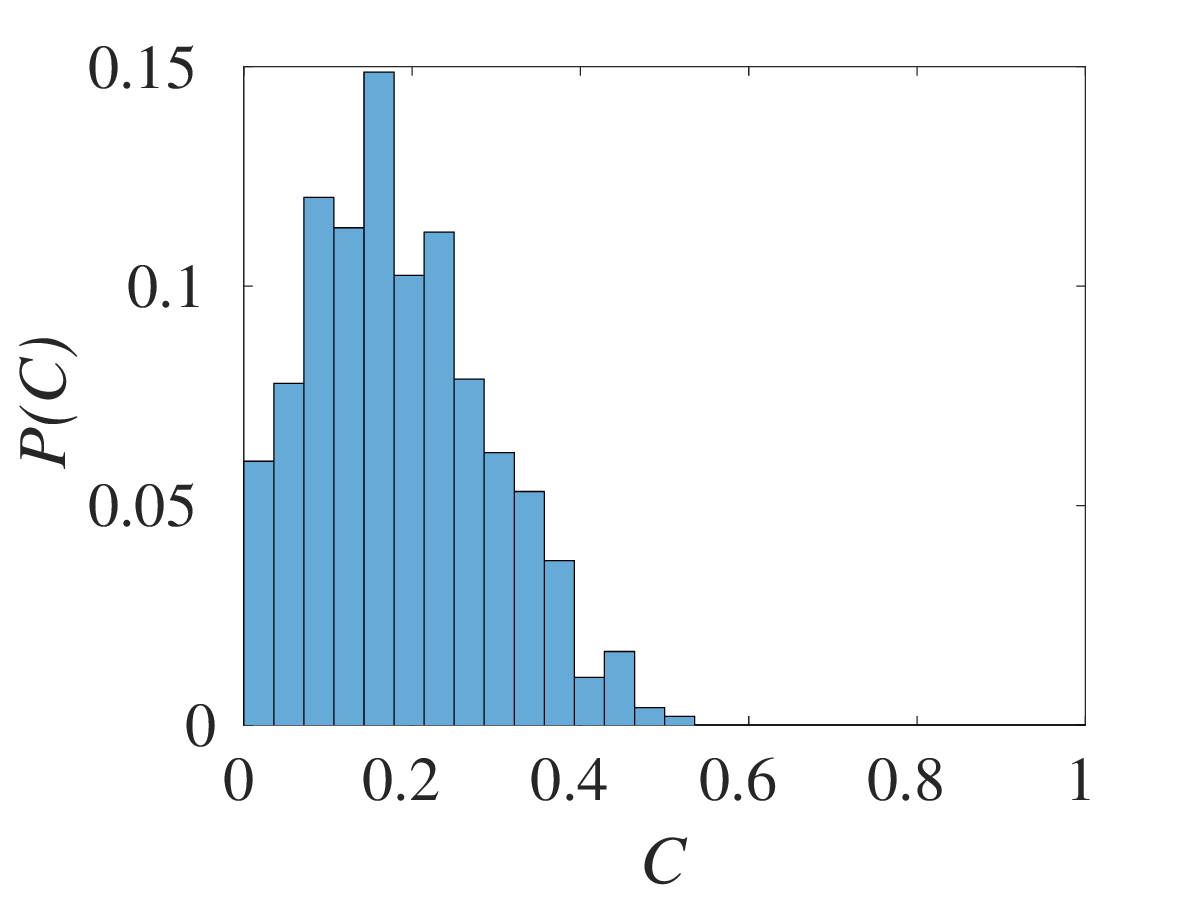}
{\bf (c)}\hskip0.45\columnwidth {\bf (d)}
\includegraphics[width=0.5\columnwidth]{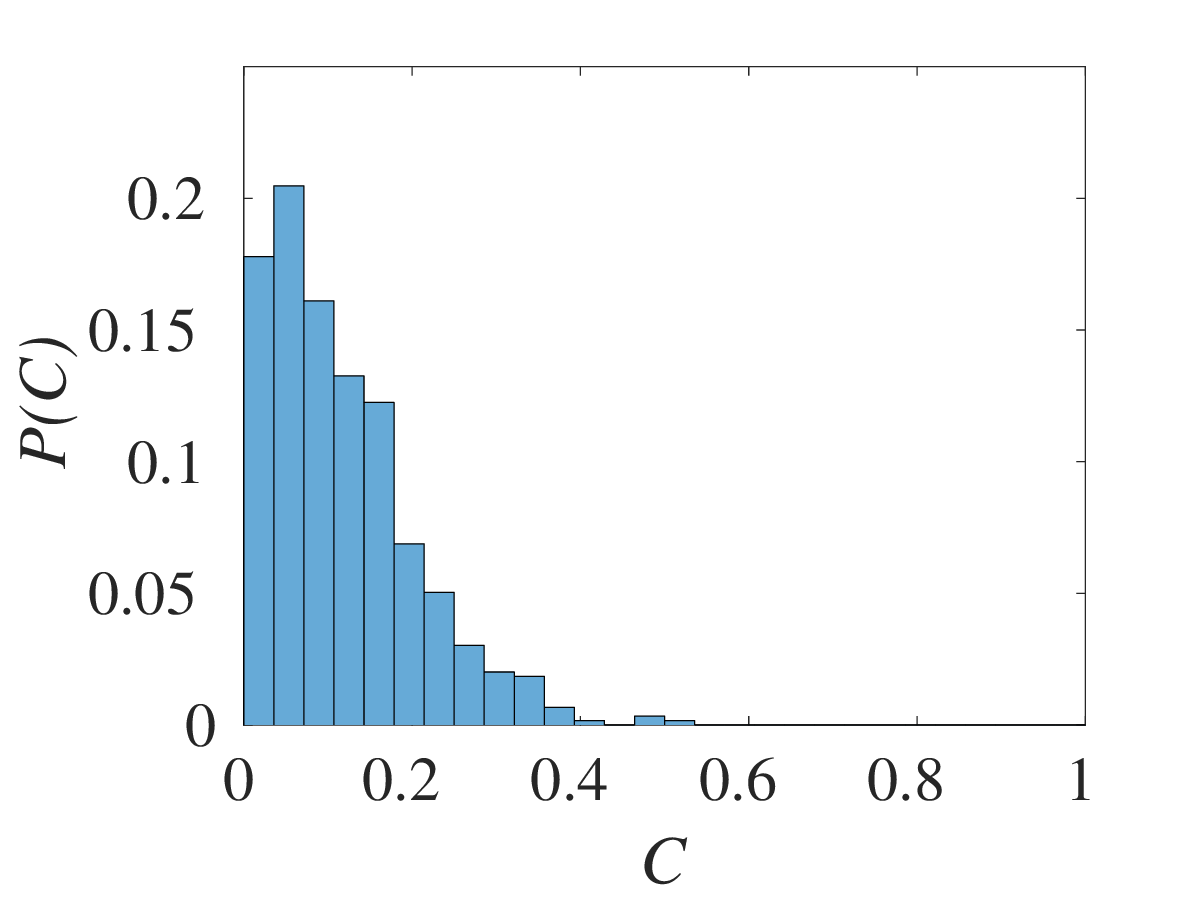}\includegraphics[width=0.5\columnwidth]{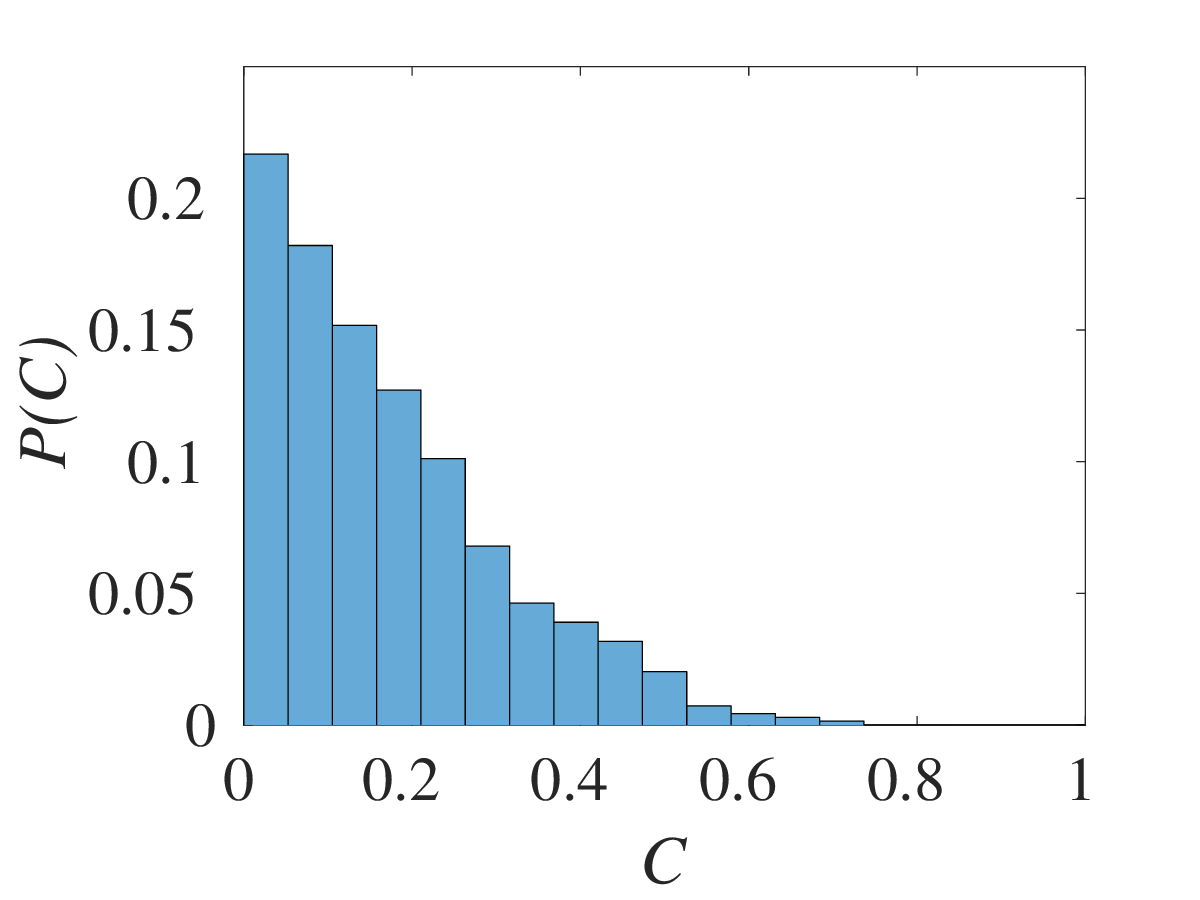}
\caption{(Color online) Long-time distribution of the concurrence. The histograms show the probability of observing a value of concurrence in each interval. The panels are for {\bf (a)} $\eta=0.5$; {\bf (b)} $\eta=1$; {\bf (c)} $\eta=2$; {\bf (d)} $\eta=4$. We have excluded the data values $C=0$ from the first bar. Calculations are shown for $L=12$.}
\label{fig:histo}
\end{figure}

Another evidence of the ergodic-MBL transition is provided by the long-time distribution of concurrence as shown in Fig.~\ref{fig:histo}. In the plots we have excluded the values $C=0$ from the first bar since, because of the ESD phenomenon, they tend to skew the distribution. The results show that for $\eta < 2$, i.e. in the ergodic phase, the distribution for low values of $C$ has a peak at a non zero value of the concurrence and then decays rapidly to zero for large values of $C$. For $\eta \ge 4$ instead, the distribution is monotonic decaying in an exponential fashion. These observations complement our previous discussion. For low values of $\eta$ some nearest-neighbor entanglement is retained leading to a large mean concurrence as in Fig.~\ref{fig5}. For larger values of $\eta$, although the mean value is comparable, the distribution is completely different. We would like to add that similar results hold for the long-time distribution of the total correlations. Although the effect is not as strong as for the concurrence, the probability distribution of the total correlations change from a skewed distribution away from the ergodic-MBL transition to an approximately Gaussian distribution near the predicted transition point (results not shown). We remark that the distribution of block entanglement is examined in Ref.~\cite{Vosk}.

\section{Conclusions}
We have examined the dynamics of correlations, encompassing both quantum and classical natures, in a many-body localized system. Using random, factorized initial states we have shown that care must be taken regarding the choice of initial state and correlation measure. In particular, despite effectively modelling the high-temperature behavior of the system, we have shown that concurrence can exhibit markedly different behaviors depending on the initial state. This is in large part due to the occurrence of (pathological) entanglement sudden death. By employing a global measure of correlations that encompasses both classical and quantum natures, we have shown that such issues can be neatly alleviated. We therefore argue that the total correlations serve as a more useful indicator in studying the dynamics. Furthermore, the total correlations highlight a clear change in the nature of the dynamics when the system is quenched into the ergodic or the MBL phase, with the latter showing oscillations in the correlations, likely related to memory effects. Finally, we assessed a different initial state composed of tensor products of singlet pairs. Our results provide important insight into the nature of the ergodic-MBL transition and highlight the care that must be taken in choosing suitable figures of merit to assess such systems. Such an observation is particularly important in the context of the recent experimental~\cite{Smith} and theoretical~\cite{IeminiArXiv} developments in studying MBL systems where the initial state is fixed.

It is important to stress that the correlation measures we consider in this paper can be measured in experiments with ultracold atoms, trapped ions and solid state implementations of spin chains. In fact the concurrence only requires two-spin correlations while, at zero temperature, total correlations require only the single spin density matrix that can be determined from the single-spin polarisation.

\acknowledgments
We are grateful to Rosario Fazio, John Goold, Fernando Iemini and A. Russomanno for helpful discussions and exchanges. The authors acknowledge support from the John Templeton Foundation (grant ID 43467), and the EU Collaborative Project TherMiQ (Grant Agreement 618074).


\begin{thebibliography}{99}
\bibitem{OsterlohNature} A. Osterloh, L. Amico, G. Falci, and R. Fazio, Nature {\bf 416}, 608 (2002)

\bibitem{Osborne} T. J. Osborne and M. A. Nielsen, Phys. Rev. A {\bf 66}, 032110 (2002).

\bibitem{FazioRMP} L. Amico, R. Fazio, A. Osterloh, and V. Vedral, Rev. Mod. Phys. {\bf 80}, 517 (2008).

\bibitem{Plastina} L. Amico, A. Osterloh, F. Plastina, R. Fazio, and G. Massimo Palma, Phys. Rev. A {\bf 69}, 022304 (2004).

\bibitem{CC} P. Calabrese and J. Cardy, J. Stat. Mech. Theor. Exp. (2005) P04010.

\bibitem{DeChiaraJSTAT} G. De Chiara, S. Montangero, P. Calabrese, and R. Fazio, J. Stat. Mech. Theor. Exp. (2006) P03001.

\bibitem{Basko} D. Basko, I. Aleiner, and B. Altshuler, Annals of Physics {\bf 321}, 1126 (2006).

\bibitem{nandkishore2015many}  R. Nandkishore and D. A. Huse, Annu. Rev. Condens. Matter Phys. {\bf 6}, 15 (2015).

\bibitem{MooreReview} R. Vasseur and J. E. Moore, J. Stat. Mech. Theor. Exp. (2016) 064010.

\bibitem{LaflorencieReview} N. Laflorencie, Physics Reports {\bf 646}, 1 (2016).

\bibitem{Serbyn1} M . Serbyn , Z. Papi\'{c}, and D. A. Abanin, Phys. Rev. Lett. {\bf 111}, 127201 (2013).

\bibitem{Serbyn2} M. Serbyn, M. Knap, S. Gopalakrishnan, Z. Papi\'{c}, N. Y. Yao, C. R. Laumann, D. A. Abanin, M. D. Lukin, and E. A. Demler, Phys. Rev. Lett. {\bf 113}, 147204 (2014).

\bibitem{Serbyn3} M . Serbyn , Z. Papi\'{c}, and D. A. Abanin, Phys. Rev. B {\bf 90}, 174302 (2014).

\bibitem{Huse4} D. A. Huse, R. Nandkishore, and V. Oganesyan, Phys. Rev. B {\bf 90}, 174202 (2014).

\bibitem{PollmanPRL} G. De Tomasi, S. Bera, J. H. Bardarson, and F. Pollman, Phys. Rev. Lett. {\bf 118}, 016804 (2017).

\bibitem{BeraPRB} S. Bera and A. Lakshminarayan, Phys. Rev. B {\bf 93}, 134204 (2016).

\bibitem{GooldPRB} J. Goold, C. Gogolin, S. R. Clark, J. Eisert, A. Scardicchio, and A. Silva, Phys. Rev. B {\bf 92}, 180202 (2015).

\bibitem{Sarandy} J. L. C. da C. Filho, A. Saguia, L. F. Santos, and M. S. Sarandy, arXiv:1705.01957

\bibitem{Schreiber} M. Schreiber, S. S. Hodgman, P. Bordia, H. P. L{\"u}schen, M. H. Fischer, R. Vosk, E. Altman, U. Schneider, and I. Bloch, Science {\bf 349}, 842 (2015).

\bibitem{Kondov} S. S. Kondov, W. R. McGehee, W. Xu, and B. DeMarco, Phys. Rev. Lett. {\bf 114}, 083002 (2015).

\bibitem{Choi} J.-y. Choi, S. Hild, J. Zeiher, P. Schau§, A. Rubio-Abadal, T. Yefsah, V. Khemani, D. A. Huse, I. Bloch, and C. Gross, Science {\bf 352}, 1547 (2016).

\bibitem{Smith} J. Smith, A. Lee, P. Richerme, B. Neyenhuis, P. W. Hess, P. Hauke, M. Heyl, D. A. Huse, and C. Monroe, Nature Phys. {\bf 12}, 907 (2016).

\bibitem{HusePRB} A. Nanduri, H. Kim, and D. A. Huse, Phys. Rev. B {\bf 90}, 064201 (2014).

\bibitem{AletPRB} D. J. Luitz, N. Laflorencie, and F. Alet, Phys. Rev. B {\bf 91}, 081103 (2015).

\bibitem{DeltaComment} It is known that to observe MBL requires $\Delta\neq0$. However the role of interaction is still not entirely clear. From preliminary calculations it seems that the behavior of concurrence and total correlations is largely unaffected. However a complete and definitive investigation about the strength of interactions goes beyond the scope of this paper.

\bibitem{EnssArXiv} T. Enss, F. Andraschko, and J. Sirker, Phys. Rev. B {\bf 95}, 045121 (2017).

\bibitem{DeLuca} A. De Luca, B. L. Altshuler, V. E. Kravtsov, and A. Scardicchio, Phys. Rev. Lett. {\bf 113}, 046806 (2014).

\bibitem{Huse5} V. Khemani, D. N. Sheng, and D. A. Huse, arXiv:1702.03932

\bibitem{ESD} T. Yu and J. H. Eberly, Science {\bf 323}, 598 (2009).

\bibitem{Vosk} R. Vosk, D. A. Huse, and E. Altman, Phys. Rev. X {\bf 5}, 031032 (2015).

\bibitem{IeminiArXiv} F. Iemini, A. Russomanno, D. Rossini, A. Scardicchio, and R. Fazio, Phys. Rev. B {\bf 94}, 214206 (2016).
\end{thebibliography}
\end{document}